\documentclass[sigconf]{acmart}

\copyrightyear{2026}
\acmYear{2026}
\setcopyright{cc}
\setcctype{by}
\acmConference[MSR '26]{23rd International Conference on Mining Software Repositories}{April 13--14, 2026}{Rio de Janeiro, Brazil}
\acmBooktitle{23rd International Conference on Mining Software Repositories (MSR '26), April 13--14, 2026, Rio de Janeiro, Brazil}
\acmPrice{}
\acmDOI{10.1145/3793302.3793597}
\acmISBN{979-8-4007-2474-9/2026/04}

\usepackage{subcaption}
\usepackage{mdframed}

\usepackage{tcolorbox}
\tcbset{
  colback=gray!15,     
  colframe=black,      
  boxrule=0.6pt,       
  arc=4pt,             
  left=6pt,            
  right=6pt,           
  top=6pt,             
  bottom=6pt           
}
\title{Novice Developers Produce Larger Review Overhead for Project Maintainers while Vibe Coding}

\author{Syed Ammar Asdaque}
\email{25280060@lums.edu.pk}
\affiliation{%
  \institution{Lahore University of Management Sciences}
  \city{Lahore}
  \state{Punjab}
  \country{Pakistan}
}
\author{Imran Haider}
\email{25280038@lums.edu.pk}
\affiliation{%
  \institution{Lahore University of Management Sciences}
  \city{Lahore}
  \state{Punjab}
  \country{Pakistan}
}
\author{Muhammad Umar Malik}
\email{27100139@lums.edu.pk}
\affiliation{%
  \institution{Lahore University of Management Sciences}
  \city{Lahore}
  \state{Punjab}
  \country{Pakistan}
}
\author{Maryam Abdul Ghafoor}
\email{maryam.ghafoor@lums.edu.pk}
\affiliation{%
  \institution{Lahore University of Management Sciences}
  \city{Lahore}
  \state{Punjab}
  \country{Pakistan}
}
\author{Abdul Ali Bangash}
\email{abdulali@lums.edu.pk}
\affiliation{%
  \institution{Lahore University of Management Sciences}
  \city{Lahore}
  \state{Punjab}
  \country{Pakistan}
}

\begin{document}
\begin{abstract}
AI coding agents allow software developers to generate code quickly, which raises a practical question for project managers and open source maintainers: can vibe coders with less development experience substitute for expert developers? To explore whether developer experience still matters in AI-assisted development, we study $22,953$ Pull Requests (PRs) from $1,719$ vibe coders in the GitHub repositories of the AIDev dataset. We split vibe coders into lower experience vibe coders ($Exp_{Low}$) and higher experience vibe coders ($Exp_{High}$) and compare contribution magnitude and PR acceptance rates across PR categories.  We find that $Exp_{Low}$ submits PRs with larger volume ($2.15\times$ more commits and $1.47\times$ more files changed) than $Exp_{High}$. 
Moreover, $Exp_{Low}$ PRs, when compared to $Exp_{High}$, receive $4.52\times$ more review comments, and have $31$\% lower acceptance rates, and remain open $5.16
\times$ longer before resolution. 
Our results indicate that low-experienced vibe coders focus on generating more code while shifting verification burden onto reviewers. For practice, project managers may not be able to safely replace experienced developers with low-experience vibe coders without increasing review capacity. Development teams should therefore combine targeted training for novices with adaptive PR review cycles.  
\end{abstract}

\maketitle

\vspace{-0.4cm}

\section{Introduction}
\label{section:introduction}

The software engineering community has long recognized developer experience and expertise as critical factors in software success~\cite{ dey2020expertise,meijer2024npm}. In traditional software engineering, seasoned developers often navigate complex codebases more efficiently, while novices focus on simpler tasks~\cite{koInfoNeeds}. Now, with the advent of ``Software $3.0$'', a paradigm in which AI agents collaborate with, or autonomously perform programming tasks~\cite{storey2025generative, Hassan2024SE3, li2025rise}, the traditional relationship between experience and task-complexity is likely changing. 
Industry reports show that most developers~\cite{sarkar2025vibecodingprogrammingconversation} already use AI coding tools ($92$\% in a survey~\cite{githubSurveyReveals}) and believe these tools help them preserve cognitive effort. Yet preliminary studies offer mixed signals: for example, a recent study~\cite{becker2025impact} measured how early-$2025$ AI tools affect experienced open-source developers during real coding tasks. The study found that developers who used AI assistants required $\sim$$19$\% more time to complete their assigned programming tasks compared to developers who did not use AI. This trend raises a key question: \textit{Does AI-based code assistance perform differently based on the coder’s experience and expertise?}

To study the patterns in vibe coding, we first establish a clear definition of the term. Prior work presents multiple interpretations of vibe coding. Maes~\cite{maes2025gotchas} characterizes it broadly as AI-assisted programming, whereas Sarkar~et~al.~\cite{sarkar2025vibecodingprogrammingconversation} and Sapkota~et~al.~\cite{Sapkota2025Agentic} offer a more precise definition: a workflow in which a human developer directs and supervises an AI agent through natural-language prompts and validates the generated code. In this study, we adopt the latter definition.
We systematically compare the contributions of vibe coders across two experience levels. We define these two groups as \textbf{$Exp_{High}$}, representing vibe coders with high development experience, and \textbf{$Exp_{Low}$}, representing vibe coders with low development experience. Following prior literature~\cite{articlemalik, articlemcIntoshKamiexpcal}, we measure a developer’s experience based on their total number of commits in the lifetime divided by their account age on GitHub.

Specifically, we study the characteristics of $22,953$ AI-based pull requests (PRs) generated by $1,719$ vibe coders on open-source Github projects. Most prior work on developer-experience research predates generative AI, and early AI-augmented studies have examined performance in isolation~\cite{storey2025generative, li2025rise}. In contrast, we investigate real-world outcomes, i.e. PR contribution magnitude (in terms of commit-frequency and number of files changed) and PR merge effort (in terms of PR acceptance rate, PR resolution time and PR review volume), under AI assistance. We ask the following research questions:\vspace{5pt}

\begin{mdframed}[backgroundcolor=gray!5,linewidth=1pt]
\noindent\textbf{RQ1:} What is the frequency and magnitude of contributions submitted by 
high-experience ($Exp_{High}$) and low-experience ($Exp_{Low}$) vibe coders in OSS?

\noindent\textbf{RQ2:} Is the PR merge effort similar between the high-experience 
($Exp_{High}$) and low-experience ($Exp_{Low}$) vibe coders?
\end{mdframed}

\vspace{5pt}


In our results, we find that, counterintuitively, $Exp_{Low}$ produce larger-scoped PRs than $Exp_{High}$, they pass the PR review phase less often and attract more reviewer
feedback, indicating that while AI enables $Exp_{Low}$ to
generate code rapidly, it fundamentally shifts the workload to
maintainers in the form of higher verification overhead. 
Thus, $Exp_{Low}$ will likely require disproportionately more development and review effort to manage these larger, slower-to-integrate
contributions. 
To support open science, we release our replication package online~\cite{replicationPackage}.
\section{Related Work}

Prior work has emphasized that developer expertise is a critical factor in software contribution. For instance, Dey et al. constructed an ``expertise space'' to model and predict developer behaviors, such as future API use and PR acceptance~\cite{dey2020expertise}. Similarly, Meijer et al. demonstrate that an individual contributor's ecosystem-wide experience (e.g., history of PRs) strongly boosts their acceptance rate and success~\cite{meijer2024npm}. These studies imply that more experienced contributors tend to make accepted, higher-quality code contributions. However, this body of work predates the widespread adoption of generative AI. Consequently, it reflects a ``Software $2.0$'' paradigm~\cite{Dilhara2021Software2, Boehm2006View20th21st, Hassan2025AgenticSE} where AI coding assistants did not mediate developer productivity. Meanwhile, the software community recognizes that generative AI represents a major paradigm shift (termed ``Software $3.0$''), fundamentally altering how developers write, debug, and understand code~\cite{li2025rise, storey2025generative}. Developers now embed AI assistants like Copilot and Cursor into their development workflows. Concurrently, surveys indicate a widespread belief that these tools will enhance productivity and allow developers to concentrate on more complex design problems~\cite{githubSurveyReveals}. Yet, controlled experiments have begun to question this optimism regarding productivity gains. For instance, Becker et al. recently found that for experienced OSS developers, using state-of-the-art AI tools actually increased task completion time by $\sim$$19$\%, suggesting new bottlenecks related to code verification or merge challenges~\cite{becker2025impact}. Li et al. reported from the AIDev dataset that AI-generated PRs tend to be accepted far less frequently than human-authored ones and are structurally simpler (fewer complexity changes)~\cite{li2025rise}. However, no prior study has systematically compared these AI-assisted contributions when authored by developers with different levels of technical experience.

\section{Dataset}
 We use the \textbf{AIDev dataset}, a publicly released corpus of AI-assisted PRs on GitHub OSS projects~\cite{li2025rise}. It includes contributions made through major AI coding agents such as Copilot, Codex, Claude Code, Cursor and Devin. {AIDev} offers a subset of PRs from repositories with more than $100$ Github stars containing $33,596$ PRs and $1,796$ users which we used for our analysis. Each record contains detailed PR-level metadata, including the number of commits, files changed, timestamps, review events, author and reviewer identifiers, and the declared PR category. These attributes provide a comprehensive foundation for analyzing contribution size, merge efficiency and review dynamics within AI-assisted development.

\section{Methodology}

We follow a three step process in our methodology. First, we select the vibe coders and their PRs. Second, we divide them into experienced groups based on their experience. Finally, we extract relevant metrics from the vibe coders' PRs for statistical analysis.

From the AIDev dataset, we select users who have created agentic PRs as \emph{vibe coders}, adopting the definition of Li~et~al.~\cite{li2025rise}. 
To rigorously isolate human activity, we filter out accounts where the username contains the sub-string \texttt{bot} or matches known agent identifiers (e.g., \texttt{Copilot}), as these represent fully autonomous agents rather than vibe coders. After filtering bots, we retain $22,953$ PRs authored by $1,719$ distinct vibe coders.

To quantify experience, we retrieve the all-time commit history for each vibe coder via the GitHub GraphQL API~\cite{GitHubGraphQL}. We aggregate contribution data across each user's entire active timeline, starting from their account creation date. Following the approach of Malik~et~al.~\cite{articlemalik}, we compute the experience score:
{\small
\[
\text{Experience Score} = 
\frac{\text{Total \# of commits by the vibe coder}}{\text{Vibe coder's account age}}
\]
} 

\noindent We distribute the $1,719$ vibe coders into four quartiles based on their experience score.
$859$ coders fall in the top two quartiles (Q$3$ \& Q$4$), whom we refer to as high-experienced developers ($Exp_{High}$) that perform vibe coding, and $860$ fall in the bottom two quartiles (Q$1$ \& Q$2$), whom we refer to as low-experienced developers ($Exp_{Low}$) that perform vibe coding.

\label{section:results}
We perform all data processing and analysis using Python (pandas, numpy, matplotlib, seaborn, and scipy). To compare the two experience groups ($Exp_{High}$ .vs. $Exp_{Low}$), we conduct group-wise statistical test comparisons on mean values of each metric across PR categories (e.g., bug fix, feature development, documentation~\cite{li2025rise}). We then interpret the observed patterns in light of the two research questions outlined in
Section~\ref{section:introduction}.
\\From each PR in the AIDev dataset, we extract four metrics:
\textbf{contribution magnitude} (number of commits and files changed per PR),
\textbf{PR acceptance rate} (merged PRs over total submitted PRs),
\textbf{PR resolution time} (days from PR creation to merge),
and \textbf{PR review volume} (total review comments per PR).

\section{Results}
In order to compare the PRs of vibe coders of $Exp_{High}$ and $Exp_{Low}$, we first analyze the frequency and magnitude of their contributions.

\subsection*{RQ1: Magnitude of Contributions ($Exp_{High}$ .vs. $Exp_{Low}$)}
\begin{figure}[t!]
    \centering
    \begin{subfigure}[b]{1\linewidth}
        \centering
        \includegraphics[width=\linewidth]{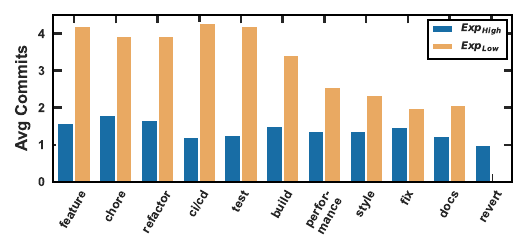}
        \vspace{-0.8cm}
        \caption{Commits per PR}
        \label{fig:commits_per_pr}
    \end{subfigure}
    \vspace{-0.6cm}
    \hfill 
    \begin{subfigure}[b]{1\linewidth}
        \centering
        \includegraphics[width=\linewidth]{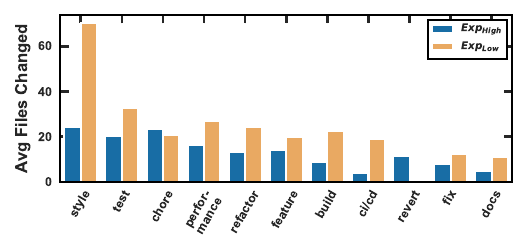}
        \vspace{-0.8cm}
        \caption{Files Changed per PR}
        \label{fig:files_changed_per_pr}
    \end{subfigure}
    \vspace{-0.1cm}
    \caption{Comparison of Number of Commits and Files Changed per Pull Request~(PR) for $Exp_{Low}$ and $Exp_{High}$.}
    \label{fig:pr_metrics}
\vspace{-0.6cm}
\end{figure}

\textbf{We find that $Exp_{Low}$ perform $2.15\times$ more AI-based commits per PR compared to $Exp_{High}$}. The Mann-Whitney U test on number of commits confirms the difference between $Exp_{Low}$ and $Exp_{High}$ as statistically significant ($p<0.05$). To determine if this pattern persists across specific PR categories, we further apply the Mann-Whitney U test to each category. Referring to \autoref{fig:commits_per_pr}, we observe that $Exp_{Low}$ performs more commits than $Exp_{High}$ in $10/11$ categories. We find that the difference remains statistically significant for all $10$ PR categories ($p<0.05$) notably including the major difference in feature related PRs ($1.58$ mean commits per PR for $Exp_{High}$ .vs. $4.20$ mean commits per PR for $Exp_{Low}$).

\textbf{We find that $Exp_{Low}$ modify $1.47\times$ more files per PR compared to $Exp_{High}$}. The Mann-Whitney U test on files changed per PR confirms the difference between $Exp_{High}$ and $Exp_{Low}$ as statistically significant ($p<0.05$). To determine if this pattern persists across specific PR categories, we further apply the Mann-Whitney U test to each category. Referring to \autoref{fig:files_changed_per_pr}, we observe that $Exp_{Low}$ modify more files than $Exp_{High}$ in $9/11$ categories. This difference is statistically significant for $5/11$ categories ($p<0.05$), notably including the major difference in style related PRs ($24.29$ mean files changed per PR for $Exp_{High}$ .vs. $70.35$ mean files changed per PR for $Exp_{Low}$).

\begin{tcolorbox}Analyzing the number of commits and changed files per PR,
we find that $Exp_{Low}$’s PRs are larger in both scope ($1.47\times$ more
files changed) and magnitude ($2.15\times$ more commits per PR).\end{tcolorbox}

\label{section:results}
\begin{figure*}[htbp!]
    \centering
    \begin{subfigure}[b]{0.32\textwidth}
        \includegraphics[width=\textwidth]{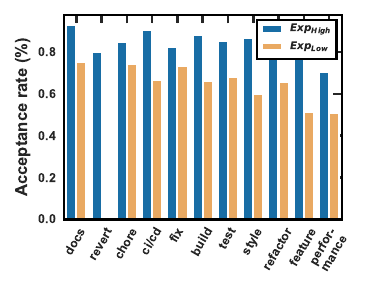}
        \caption{PR Acceptance Rate}
        \label{fig:par}
    \end{subfigure}
    \hfill
    \begin{subfigure}[b]{0.32\textwidth}
        \includegraphics[width=\textwidth]{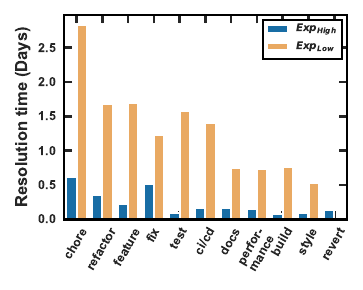}
        \caption{PR Resolution Time}
        \label{fig:resolution_time}
    \end{subfigure}
    \hfill
    \begin{subfigure}[b]{0.32\textwidth}
        \includegraphics[width=\textwidth]{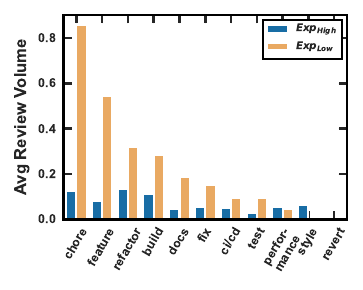}
        \caption{PR Review Volume}
        \label{fig:review volume}
    \end{subfigure}
    
    \caption{Comparison of PR acceptance rate, PR resolution time and PR review volume for $Exp_{Low}$ and $Exp_{High}$.}
    \label{fig:dev_metrics_resolution}
\end{figure*}

\subsection*{RQ2: PR Merge Effort($Exp_{High}$ .vs. $Exp_{Low}$)} 

\autoref{fig:dev_metrics_resolution} shows the comparison between $Exp_{High}$ and $Exp_{Low}$ across $11$ categories based on PR acceptance rate, PR resolution time, and PR review volume.

\textbf{We find that $Exp_{Low}$ PRs have $31\%$ lower PR acceptance rate as compared to $Exp_{High}$}. We confirm using a Chi-Square test~\cite{pearson1900criterion} that this difference is significant, $p<0.05$. 
To determine if this pattern persists across specific PR categories, we further apply the Chi-Square test to each category. Referring to \autoref{fig:par}, we observe that $Exp_{Low}$ have lower PR acceptance rate than $Exp_{High}$ in $10/11$ categories. We find a significant difference in all of these $10$ PR categories ($p<0.05$), notably including the major difference in documentation related PRs ($93.06\%$ acceptance rate for $Exp_{High}$ .vs. $75.39\%$ acceptance rate for $Exp_{Low}$).

Next, we measure mean PR resolution time (elapsed time from PR creation to merge), as shown in \autoref{fig:resolution_time}. \textbf{We find that $Exp_{Low}$ takes $5.16\times$ more time to resolve PRs than $Exp_{High}$.}
The difference between the two experience groups is significant, as we confirm through a Mann-Whitney U test~\cite{mann1947test}, $p<0.05$. 
On category level analysis, using Mann-Whitney U test~\cite{mann1947test}, we find statistically significant difference in $10/11$ PR categories, including the major difference in chore related PRs ($0.61$ days for $Exp_{High}$ .vs. $2.83$ days for $Exp_{Low}$).

To diagnose the reasons for longer PR resolution time in $Exp_{Low}$, we compare the PR review volume of the two experience groups as shown in \autoref{fig:review volume}. 
PR review volume is the total number of review comments received per agentic PR.
\textbf{We find that $Exp_{Low}$ PRs receive $4.52\times$ more review comments than $Exp_{High}$ PRs.}
The difference between the two experience groups is significant, as we confirm through a Mann-Whitney U test~\cite{mann1947test}, $p<0.05$. 
On category level analysis, using Mann-Whitney U test~\cite{mann1947test}, we find statistically significant difference in $6/11$ PR categories, including the major difference in chore related PRs (mean $0.13$ review comments per PR for $Exp_{High}$ .vs. mean $0.86$ review comments per PR for $Exp_{Low}$).

\begin{tcolorbox}
A lower PR acceptance rate for $Exp_{Low}$ than $Exp_{High}$ indicates that novice vibe coders 
face the steepest barrier to PR acceptance. While Li~et~al.~\cite{li2025rise} 
previously observed that AI-generated PRs generally face lower acceptance 
rates, our results refine this understanding: the struggle to achieve 
mergeable quality PR is more prevalent in $Exp_{Low}$. This result suggests that despite the 
aid of \emph{vibe coding}, novice vibe coders may still lack the requisite 
expertise to \textbf{verify} and \textbf{tune} AI generated code to a standard that would satisfy OSS reviewers.
\end{tcolorbox}

\section{Potential Struggles of Novice Vibe Coders}
\label{section:discussion}
We analyze $22,953$ PRs from vibe coders to compare
less-experienced vibe coders with more experienced
vibe coders. $Exp_{Low}$ make $2.15\times$ larger and $1.47\times$ broader
changes, as reflected in higher commits and changed files per PR,
but reviewers spent $5.16\times$ more time evaluating and resolving
$Exp_{Low}$ PRs. $31\%$ lower PR acceptance rate and $4.52\times$ higher
PR review volume indicate that reviewers had to provide more feedback, identify issues, and guide corrections for PRs from less-experienced vibe coders. To understand these frictions, we manually inspect the top $15$ highest review volume PRs in the feature development category, which is the most frequent type of PR in our dataset, submitted by $Exp_{Low}$. As a result, we identify two potential sources of friction:\\\textbf{1. Infrastructure Mismatch:} $Exp_{Low}$ vibe coders struggle to align AI-generated logic with the constraints of the build environment. We observe a recurring pattern where AI models generate syntactically correct code but often fail to account for environment-specific runtime factors. To illustrate this broader trend, we highlight a case in the roboflow/inference repository's PR\#$1350$~\cite{githubPerceptionEncoder}, where the vibe coder faces significant friction dealing with execution timeouts in a CI environment. The PR illustrates a struggle not with business logic but with the limitations of the infrastructure. The vibe coder has to commit repeatedly to tune timeout parameters, effectively debugging via CI because the vibe coder likely could not replicate the constrained environment locally.\\
\textbf{2. Integration Friction:} We observed that AI-generated code often lacked necessary system context, resulting in extended review cycles for $Exp_{Low}$ authors. For instance, in getsentry/sentry repository's PR\#$94889$~\cite{githubFeatuserFeedback}, the author implemented user feedback mechanisms but faced challenges aligning the code with the repository's existing privacy schemas and integration standards. The high review volume indicates that while the feature logic was generated, ensuring compliance with project-specific architectural requirements required significant manual verification and adjustment.

\section{Implications}
\textbf{Project managers and maintainers} should anticipate higher reviewer workload and longer PR resolution times by $Exp_{Low}$ vibe coders. They can mitigate this challenge by integrating efficient review strategies, such as assigning extra reviewers or implementing automated review checks for $Exp_{Low}$ PRs.\\\textbf{Training programs and community onboarding} must emphasize verification skills for AI-generated code, equipping $Exp_{Low}$ vibe coders to evaluate it for correctness, style, and security. Finally, the $Exp_{High}$ .vs. $Exp_{Low}$ analysis offers a new analytical lens for studying AI-augmented software development. Extending our approach to industrial settings or longitudinal studies could reveal how AI’s role evolves with experience and provide an empirical foundation for designing adaptive AI tools, reviewer automation strategies, and task-assignment models in the era of Software $3.0$.

\section{Threats to Validity}
\label{section:validity}

The definition of vibe coding varies from broad AI assistance~\cite{maes2025gotchas} to strict agent supervision via prompting and verification~\cite{sarkar2025vibecodingprogrammingconversation,Sapkota2025Agentic}. Our findings may not apply to contexts using the broader definition of vibe coding. Second, we measure ``experience'' via commit count \cite{articlemalik}, which conflates activity with skill. A developer with broad offline experience but few GitHub contributions would count as ``low-experienced'' in our work. However, in the context of AI workflows (where familiarity with agents matters),  experience is a reasonable proxy. Third, project-specific factors (e.g., review policies) can influence PR acceptance rate and PR review volume. We mitigate this bias by comparing means of metrics within categories, but residual effects may remain. Fourth, conducting multiple statistical comparisons increases the risk of Type I errors (false positives). To mitigate this, we performed Benjamini-Hochberg (BH) correction~\cite{benjamini1995controlling} for all tests, ensuring that reported significant differences are robust.

\section{Ethical Considerations}
The study did not require ethical approval because it did not involve human participants. We used the AIDev dataset and cited its authors throughout the paper to give appropriate credit. We removed GitHub repository author names for their privacy.

\section{Conclusion}
\label{section:conclusion}
This study addresses whether novice developers who engage in vibe coding can substitute for experienced developers who also engage in vibe coding. Our study indicates that while vibe coding enables novices ($Exp_{Low}$) to produce much larger contributions, it also incurs substantial verification cost. In the AIDev dataset, $Exp_{Low}$ PRs exhibit $2.15\times$ higher code magnitude, measured by commits, but also face $5.16\times$ longer PR resolution time. $31\%$ lower PR acceptance rate and $4.52\times$ higher PR review volume accompany this increase, indicating that verifying $Exp_{Low}$ generated code may demand a tremendous increase in reviewer effort. Our study provides actionable guidance for software development teams. Teams must rebalance oversight and developers' training for different experience levels by implementing mandatory code reviews or monitoring review volume spikes to allocate review effort effectively. Training must focus on teaching novice developers how to rigorously test and verify code outputs while vibe coding. Finally, the $Exp_{High}$ vs. $Exp_{Low}$ experience segmentation offers a robust methodological framework, providing researchers a lens to study the evolving dynamics of human-AI collaboration across various software engineering tasks.

\bibliographystyle{ACM-Reference-Format}
\bibliography{references}

\end{document}